\documentclass[a4paper,fleqn]{cas-dc}
\usepackage[authoryear,longnamesfirst]{natbib}
\usepackage{float}
\usepackage{subcaption}

\begin{document}
\let\WriteBookmarks\relax
\def\floatpagepagefraction{1}
\def\textpagefraction{.001}
\shorttitle{Ensemble Mesoscale Flame Regime Classification}
\shortauthors{Ganesh et al.}

\title [mode = title]{Machine Learning based Ensemble Flame Regime Classification for Mesoscale Combustors based on Insights from Linear and Nonlinear Dynamic Analysis}                      
\author[1]{M Ashwin Ganesh}
\credit{Conceptualization, Data Analysis and Visualization, Machine Learning Implementation, Writing - Original Draft, Writing - Review \& Editing}

\author[1]{Akhil Aravind}
\credit{Conceptualization, Experiments and Data Acquisition, Writing - Review \& Editing}

\author[1]{Balasundaram Mohan}
\credit{Conceptualization, Experiments and Data Acquisition}

\author[1]{Saptarshi Basu} 
\cormark[1]
\ead{sbasu@iisc.ac.in}

\cortext[cor1]{Corresponding author}
\affiliation[1]{organization={Department of Mechanical Engineering, Indian Institute of Science},
                city={Bengaluru},
                state={Karnataka},
                country={India}}
\credit{Conceptualization, Funding Acquisition \& Resources, Writing - Review \& Editing,}
\begin{abstract}
Gaining insights into flame behavior at small scales can lead to improvements in the efficiency of micro-reactors, compact power generation systems, fire safety technologies, and various other applications where combustion is confined to micro or mesoscales. Flame regimes observed in mesoscale combustors, namely Stable flame, Flames with repetitive extinction and ignition, and Propagating flame, exhibit unique dynamic characteristics that differentiate them from one another. In this study, we systematically examine the various flame regimes observed in mesoscale combustors from both dynamical and statistical standpoints. Our experimental methodology involves stabilizing a flame inside a quartz tube (an optically accessible mesoscale combustor) with an inner diameter of 5 mm. A premixed methane-air mixture is used as fuel, with its equivalence ratio and Reynolds number being the input parameters. Instantaneous OH$^*$ chemiluminescence and Acoustic pressure signals, along with high-speed flame imaging, were acquired for combustion dynamics characterization. The objective of this study is to analyze the distinct dynamical signatures associated with these observed flame regimes. For this purpose, Recurrence Quantification Analysis, followed by a Statistical-Spectral analysis, has been performed based on the experimentally acquired OH$^*$ Chemiluminescence and Acoustic pressure time-series signals. Subsequently, a stacking ensemble-based machine learning framework has been implemented for mesoscale flame regime classification based on the features extracted from the two aforementioned analyses. In addition, Continuous wavelet transform (CWT) scalograms and three-dimensional phase plots have been graphed to visually elucidate the evolution of system dynamics and the complex interaction of competing time scales in these flame regimes. Overall, this study aims to provide an insightful understanding of the various flame regimes observed in mesoscale combustors, integrated with a stacking ensemble framework intended for flame regime classification. 
\end{abstract}

\begin{graphicalabstract}
\includegraphics[width=\linewidth]{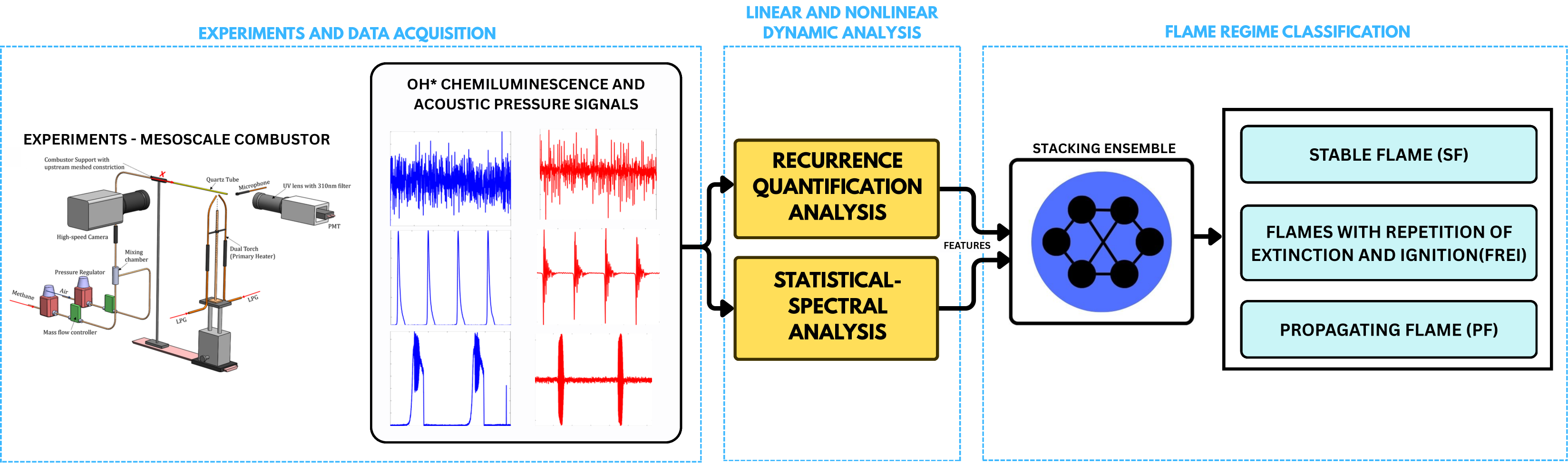}
\end{graphicalabstract}

\begin{highlights}
\item Mesoscale combustor flame regimes analyzed from both dynamical systems and statistical standpoints.
\item Recurrence Quantification and Statistical-Spectral analyses on experimentally acquired signals revealed interesting physical insights into their evolution dynamics.
\item Recurrence plots, Phase plots and Scalograms visually elucidated the dynamical signatures unique to each flame regime.
\item A Stacking ensemble framework implemented for flame regime classification achieved high accuracy.
\item Isomap and PCA demonstrated distinct clustering in the acquired feature spaces.
\end{highlights}


\maketitle

\section{Introduction}

Combustion is a primary competitor in replacing batteries in micro-electro-mechanical systems (MEMS) owing to their high energy density, rapid recharge time, and compact size compared to conventional batteries \cite{ju2011microscale}. Significant developments in this field have improved our understanding of micro-power generators (MPGs) over the last few decades \cite{Aravind2011_ApplEnergy, Aravind2013_ApplEnergy}. In MPGs, combustion takes place in narrow channels, measuring $\mathcal{O}(10^0)$ mm or smaller. The channel width is comparable to the quenching distance associated with hydrocarbon fuels. The distinction between micro- and mesoscale combustion is based on the relative size of the combustor diameter compared to the quenching distance. In micro-combustors, the characteristic combustor dimension is smaller than the quenching diameter, whereas in mesoscale combustors, it exceeds the quenching diameter but remains in the same order of magnitude. Combustion within these confined channels is strongly influenced by flame-wall interactions, leading to heat losses that can cause thermal quenching or surface reactions with the combustor walls that result in radical quenching. These processes play a critical role in shaping the flame structure, stability, and overall combustion characteristics \cite{ju2011microscale}.

Practical micro/mesoscale combustors employ product gas upstream recirculation along the combustor walls to preheat the reactant and modify the wall temperature profile to prevent thermal and radical quenching. However, this approach introduces additional flame regimes that are otherwise not observed in large-scale combustors. In addition to stationary steady flames, Maruta et al. \cite{maruta_characteristics_2005} observed unsteady flames exhibiting cycles of ignition, propagation, extinction, and re-ignition—referred to as FREI (Flames with Repetitive Extinction and Ignition)—as well as pulsating flames and flames with characteristics of both pulsation and FREI, in a 2 mm cylindrical quartz tube acting as an optically accessible micro combustor. Fan et al. \cite{fan_experimental_2009}, and Richecoeur et al. \cite{richecoeur_experimental_2005} reported similar unsteady flame behaviors in rectangular and curved mesoscale channels, respectively. 
Additionally, studies (\cite{richecoeur_experimental_2005, mohan2023self}) have also reported acoustic emission corresponding to the observed unsteady flame regimes. 
In the studies reported by Nicoud et al. and David et al. \cite{nicoud_thermoacoustic_2005, d_p_experimental_2012}, the authors attributed the sound generation to the contraction/expansion that happens during the extinction/re-ignition events associated with unsteady flames. However, recent studies by Aravind et al. \cite{aravind2023dynamics} have reported an additional unsteady flame regime in lean premixed methane-air mixtures, termed the propagating flame, wherein thermo-acoustic coupling was observed in a periodically repeating flame exhibiting cycles of ignition, propagation, and extinction.

The Recurrence Quantification Analysis (RQA) technique is widely used for nonlinear time-series and complex dynamical systems analysis, quantifying system dynamics by capturing repetition, laminarity, determinism, and stochasticity through the distinct geometric structures observed in recurrence plots. Recurrence plots were introduced as graphical tools to visualize the recurrence of states in dynamical systems as a two-dimensional map, wherein distinct patterns represent specific dynamical regimes of the system \cite{eckmann1987recurrence}. To facilitate the transition from qualitative visualization to quantitative analysis, Zbilut et al. introduced Recurrence Quantification Analysis (RQA) \cite{zbilut1992embeddings}, defining recurrence-based features to objectively characterize the dynamical state of the system. Addressing the limitation of certain preliminary RQA metrics, which only analyzed diagonal lines, Marwan et al. \cite{marwan2002recurrence} introduced measures based on vertical line structures to capture laminar states and intermittency.  Finally, Marwan et al. \cite{marwan2007recurrence} consolidated the field by formalizing Cross and Joint Recurrence Analysis and establishing rigorous best practices for phase space reconstruction.

Integrating Recurrence Quantification Analysis (RQA) with Machine Learning (ML) algorithms for anomaly detection has attracted significant research interest across diverse scientific disciplines. In the context of combustion, this approach has been majorly exploited to predict thermoacoustic instabilities. Foundational studies by Sujith et al. \cite{godavarthi2017recurrence} and Gotoda et al.\cite{gotoda2011dynamic}  first established the utility of recurrence networks and symbolic dynamics for characterizing the transition from combustion noise to instability. To address the critical issue of flame extinction in gas turbine engines, De et al. \cite{de2020application}  utilized CH* chemiluminescence time series from a swirl-stabilized dump combustor, identifying specific recurrence parameters as robust precursors for lean blowout. Waxenegger-Wilfing et al. \cite{waxenegger2021early} performed RQA on dynamic pressure time-series data from a cryogenic LOX/H2 rocket combustor, utilizing them to train a Support Vector Machine model to classify combustion regimes to predict the onset of high-frequency thermoacoustic instabilities. Kang et al. \cite{kang2025graph} proposed a novel Graph Convolutional Neural Network-based architecture to explicitly model the non-linear recurrence relationships embedded in the data, leading to a more accurate and efficient classification of multivariate time series. Zeng et al. \cite{zeng2024monitoring} characterized the non-linear transition from stable combustion to thermoacoustic oscillation by integrating RQA measures with an optimized Deep Belief Network. In addition, a few others have explored anomaly detection in combustion using machine learning architectures such as Bayesian Networks, Convolutional Autoencoders, etc. \cite{Sengupta2023ForecastingThermoacoustic, Maeta2024ImprovingAccuracyCombustionInstability}. These, along with several other ongoing developments, highlight the potential of combining RQA features with ML algorithms to develop a more robust framework for predicting combustion instabilities.

Building upon the aforementioned studies, this work investigates the dynamics of mesoscale flame regimes using linear \& nonlinear time-series analysis, followed by a stacking ensemble framework for flame regime classification based on the acquired features. Non-linear time series analysis is widely employed in thermoacoustics to successfully elucidate the underlying states such as high-frequency noise, periodic cycles, quasi-periodicity, chaos, and intermittency \cite{juniper2018sensitivity}. These tools and their measures were also subsequently used as a precursor in warning of the onset of different dynamical states \cite{nair2014multifractality}. For instance, the root mean square value, signal-to-noise ratio, Hurst exponent, permutation entropy, fractal dimension, and recurrence measures are some quantities that forewarn the different states involved in gas turbine combustors \cite{kabiraj2015recurrence,vishnoi2024effect,archhith2024combustion}. Similar to the large-scale systems, the mesoscale flame regimes investigated in this study have characteristic acoustic and heat release rate signatures, which could be used to classify the various flame regimes using the above-mentioned tools \cite{aravind2023dynamics,archhith2024combustion}. To the extent of the author's knowledge, these tools were not explicitly applied in the context of mesoscale combustors. In this article, we attempt for the first time to apply some of these linear and nonlinear time series analysis tools to get insight into the various flame regimes and elucidate the physics behind their topologies, followed by a stacking ensemble framework for the flame regime classification. As part of this study, Recurrence Quantification Analysis, followed by Statistical-Spectral Analysis, has been performed on experimentally acquired OH* Chemiluminescence and Acoustic pressure time-series signals, which were subsequently provided independently as input to a stacking ensemble framework to interpret the unique distinguishable dynamical signatures associated with each flame regime.

The major contributions of this work include (i) Recurrence Quantification and Statistical \& Spectral Analysis to gain insights into various flame regimes observed in mesoscale combustors and thereby elucidate the physics behind their topologies. (ii) A stacking ensemble framework for ternary flame regime classification based on the features obtained from the aforementioned analyses, for mesoscale combustors.

This article has been structured as follows. Firstly, we briefly discuss the experimental setup and data acquisition processes. Next, we explain the preprocessing involved, linear \& non-linear time series analysis tools, and the machine learning framework employed. Subsequently, a characterization of various flame regimes has been presented. Finally, the results based on flame regime classification are interpreted, followed by a discussion of the results and concluding remarks. 

\section{Experimental Setup and Instrumentation\label{Exp_setup}} \addvspace{10pt}

\begin{figure}[pos=H]
    \centering
    \includegraphics[width=\linewidth]{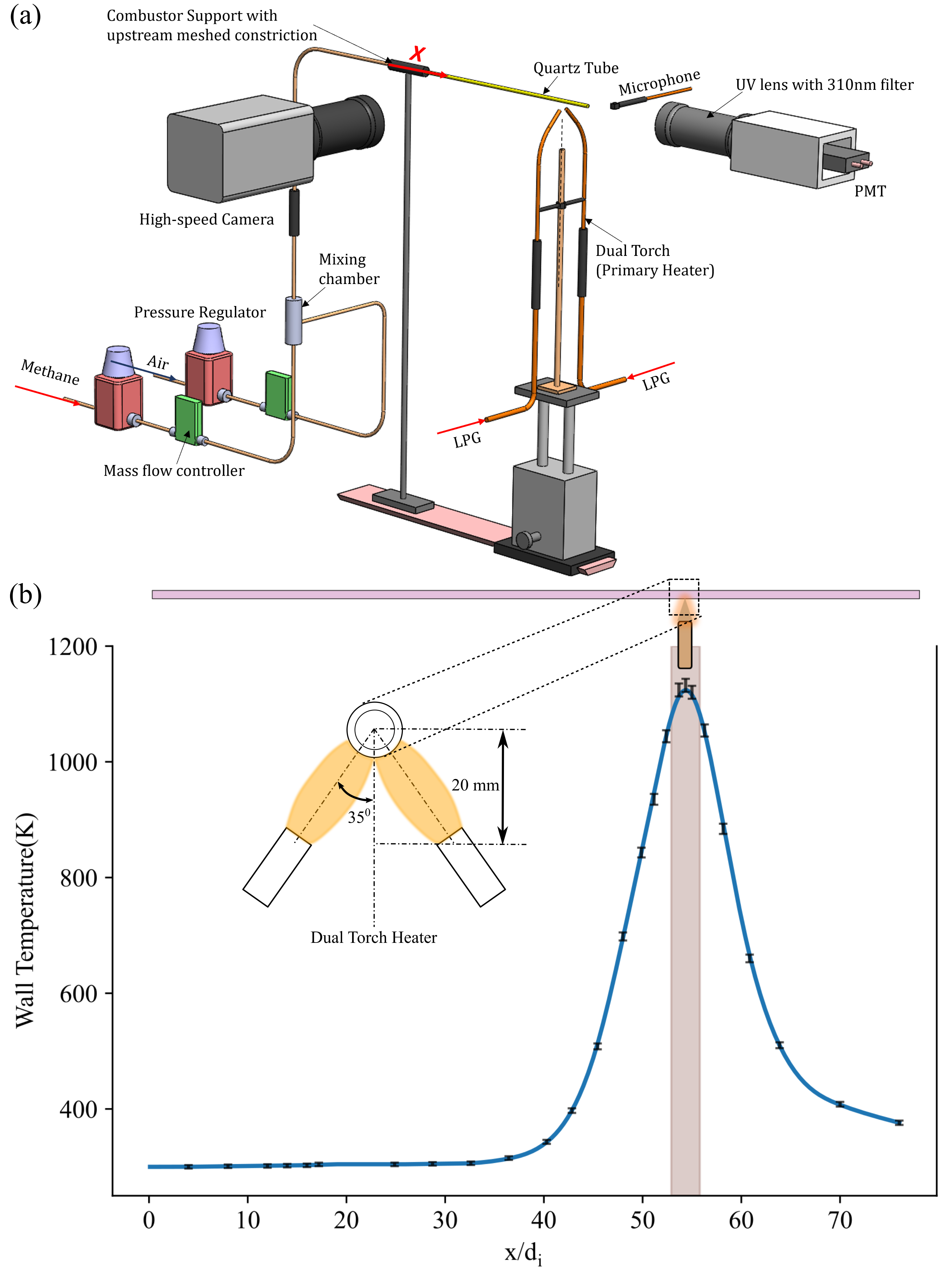}
    \caption{(a) Schematic of the experimental setup with flow lines, diagnostic tools, and the external heater \cite{aravind2023dynamics}. (b) Inner wall temperature profile plotted along the combustor axis.}
    \label{fig:Exp_Setup}
\end{figure}

The experimental facility consists of a cylindrical quartz tube serving as an optically accessible mesoscale combustor, along with an external heater that preheats the reactants (simulating the effect of upstream heat recirculation in practical mesoscale combustors) before they reach the reaction zone. The quartz tube is 320 mm long and has an inner diameter of 5 mm and an outer diameter of 7 mm. In the discussions that follow, the $x$-axis is oriented along the combustor axis, extending along the downstream direction. The origin ($x=0$ mm) is set at the upstream end of the tube. The quartz tube connects to the upstream mixing chamber via a $1.5$ mm tubular constriction followed by a flashback arrestor. The $1.5$ mm tube houses a $100 \mu m$ wire mesh upstream of $x=0$ mm. The external heater consists of two LPG flame torches arranged at an angle and is positioned 273 mm from the upstream end of the quartz combustor tube (Fig. \ref{fig:Exp_Setup}(a)). The inner wall temperature profile of the quartz tube, measured using a 1 mm K-type thermocouple, is shown in Figure \ref{fig:Exp_Setup}(b). The profile exhibits a sharp rise and fall in combustor wall temperature along the tube length, reaching a peak of approximately $\sim1130 K$ at the location corresponding to the external heater.

Methane and air regulated through two precise mass flow controllers (Bronkhorst Flexi-Flow Compact with the range of 0-1.6 SLPM for CH$_4$ and 0-2 SLPM for air) are directed into a mixing chamber as shown in Figure \ref{fig:Exp_Setup}(a). These streams mix into each other to create a homogeneous mixture, which is subsequently fed into the quartz combustor tube at its upstream end. In this study, we report measurements corresponding to three different equivalence ratios ($\Phi$): $0.8$, $1.0$, and $1.2$ with the mixture velocity ($\bar{u}$) varied between $0.2-0.7$ m/s with an increment of $0.1$ m/s. This leads to upstream Reynolds number (Re) variation of $64-224$ in increments of $32$. The flow rates of LPG (liquefied petroleum gas) and air into the external heater are controlled using precise pressure regulators and mass flow controllers (Alicat Scientific MCR-500SLPM), respectively.

\begin{figure}[pos=H]
    \centering
    \includegraphics[width=\linewidth]{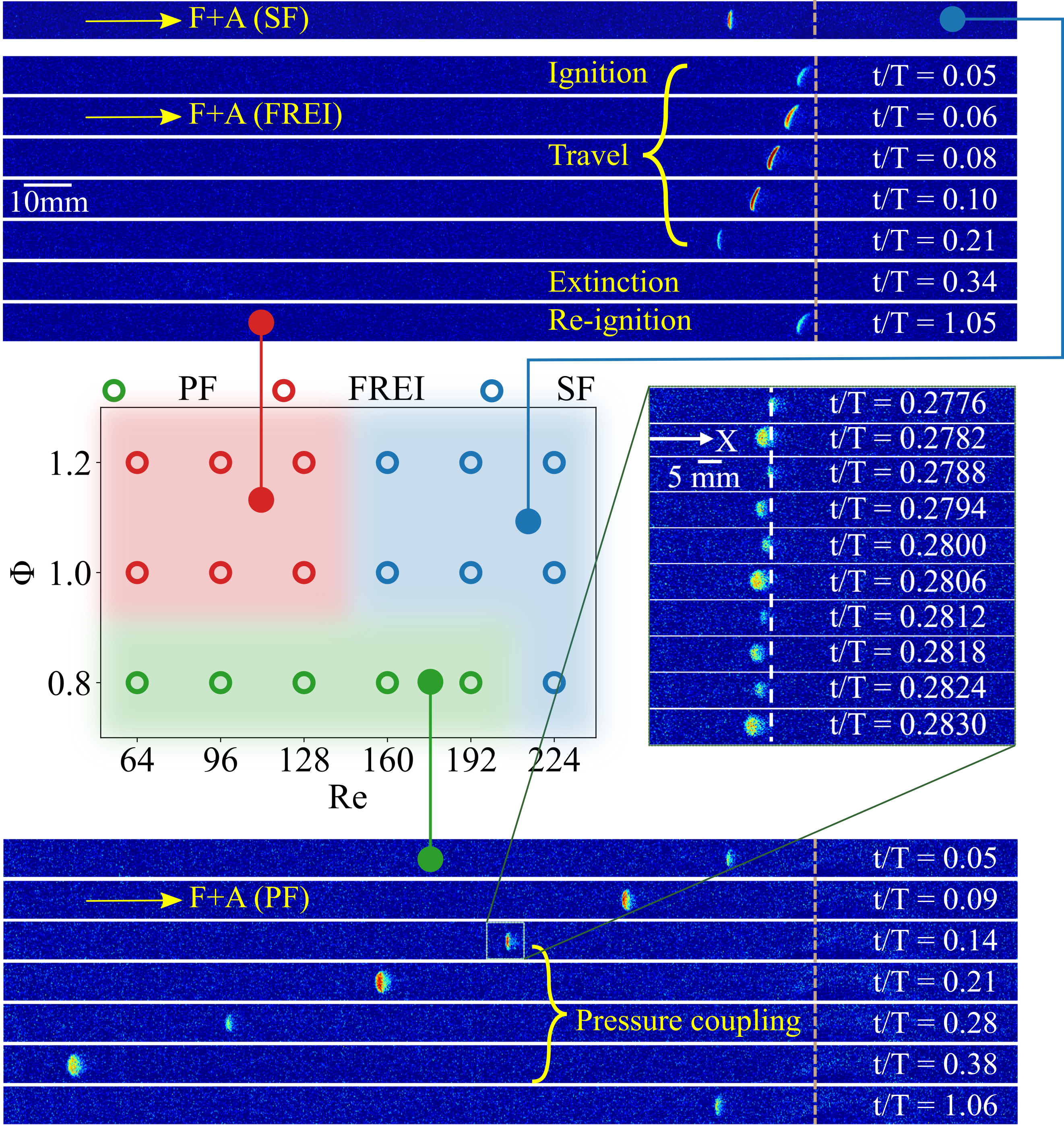}
    \caption{The Flame regime map for Re versus $\Phi$. The blue, red, and green colors on the map denote Stable Flame (SF), Flames with Repetition of Extinction and Ignition (FREI), and Propagating Flame (PF) regimes, respectively.}
    \label{fig:Regime_Map}
\end{figure}

A Phantom Miro Lab 110 high-speed camera coupled with a $100$ mm Tokina macro-lens was used for high-speed flame imaging. The dynamics were captured at 4000 frames per second ($250 \mu s$ exposure time) with a spatial resolution of $200 \mu m$ per pixel (frame size of 1280 x 120 pixels). The data was used to track the spatial location of the flame. Representative time series of the flame images are presented in Figure \ref{fig:Regime_Map}. The readers are referred to the work of Aravind et al. \cite{aravind2023dynamics} for a more detailed depiction of the flame images and their associated measures. The OH* chemiluminescence signal of the flame was captured using a Hamamatsu photomultiplier tube (PMT-H11526-110-NF). The PMT was positioned at a distance of $70$ mm from the downstream end of the combustor along the tube axis. The photocathode is exposed to the flame inside the combustor tube via a Nikon Rayfact (PF10445MF) UV lens and an OH* bandpass filter ($310\pm10$ nm) as depicted in Figure \ref{fig:Exp_Setup}(a). Since the flame is premixed, the OH$^*$ chemiluminescence signal is directly related to the heat release rate. The pressure fluctuations were recorded using a PCB microphone (PCB 130E20), which was placed 40mm downstream of the combustor exit. The data from the PMT and microphone were acquired using an NI-DAQ (PCI 6251) at 12 kHz and were triggered alongside flame imaging via the high-speed camera.

\section{Methodology}

\subsection{Signal Preprocessing}
OH* Chemiluminescence and Acoustic Pressure signals experimentally acquired at every operating condition were preprocessed in the following manner before the Recurrence quantification and Statistical-Spectral analysis. A sliding segment approach has been employed, with a defined overlap, with segment and step sizes selected appropriately to ensure the resulting subsequences of each signal statistically preserve the underlying dynamics of their corresponding complete time series \footnote{Comparison of probability distributions for segmented and global signals has been provided with the Supplementary material for interested readers.}. Furthermore, this sliding segment formulation functioned as an implicit data augmentation strategy, substantially expanding the effective dataset size through this sampling while capturing the underlying dynamic signal characteristics. Z-score normalization was applied to each signal segment to standardize variance, ensuring that the reconstructed attractors share a consistent scale. Thereby, these preprocessed signals physically imply the fluctuations in OH* chemiluminescence intensity and acoustic pressure about their respective mean.

\subsection{Recurrence Quantification Analysis}
\setlength{\abovedisplayskip}{8pt}
\setlength{\belowdisplayskip}{8pt}

Phase space reconstruction has been performed, followed by evaluation of  RQA measures to gain insight into the nature of fluctuations associated with different flame regimes in the mesoscale combustor \cite{yang2011multiscale,meyers2020cross}.  Phase space reconstruction leverages Takens' time delay embedding theorem \cite{takens1981detecting} to topologically recover a system's underlying multi-dimensional dynamics from a single univariate time series. Takens' theorem established that time-delayed observations implicitly encode the system's higher-order derivatives, enabling the reconstruction of the phase space dynamics via the delay-coordinate vector $\mathbf{Z}(i)$:
\begin{equation}
\mathbf{Z}(i) = \left[ x(i), \, x(i+\tau), \, x(i+2\tau), \, \ldots, \, x(i+(d-1)\tau) \right]
\end{equation}
where x represents the scalar time series of length $N$, $\tau$ denotes the reconstruction delay, and $d$ is the embedding dimension, defined for the range $i = 1, \ldots, N - (d - 1)\tau$.

\begin{figure}[pos=H]
    \centering
    \includegraphics[width=0.85\linewidth]{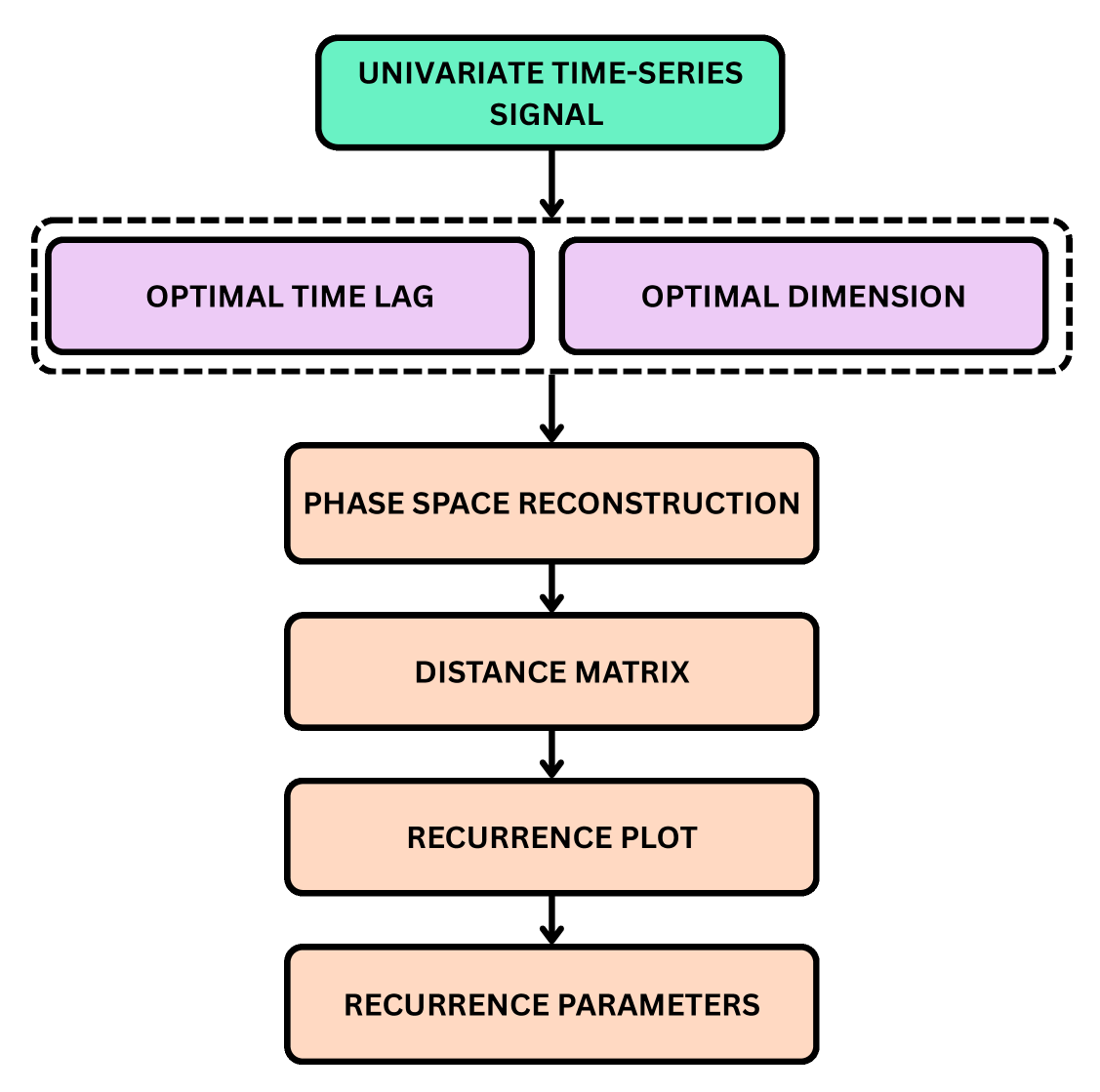}
    \caption{The workflow for Recurrence Quantification Analysis.}
    \label{RQA}
\end{figure}

Further, in order to reconstruct the phase space \cite{juniper2018sensitivity}  using these two experimental univariate time-series signals, there were two parameters to be determined, namely optimal time lag (tau) and optimal embedding dimension (d). The optimal time lag was selected as the first local minimum of the Average Mutual Information (AMI) function, while the optimal embedding dimension was identified using Cao's method to ensure the attractor was fully unfolded to demonstrate the underlying dynamics accurately. Subsequently, Recurrence Distance matrices were constructed by computing the pairwise Euclidean distances between state vectors in the reconstructed phase space. In order to ensure comparable recurrence densities, the reconstructed phase space distance matrices were thresholded using a Fixed Amount of Neighbors (FAN) scheme with a constant recurrence rate of 2\% at every state, thereby consistently capturing dominant dynamical structures across all regimes. This process workflow for recurrence quantification analysis has been sketched for clear interpretation, in Figure \ref{RQA}.

The $j^{th}$ vector reconstructed from the time series can be expressed as:
\begin{equation}
    y_j=(p^\prime(j),p^\prime(j+\tau),...,p^\prime(j+(d-1)\tau))
\end{equation}
The resulting Distance matrix can be expressed as follows:
\begin{equation}
    D_{m,n}=||y_m-y_n||
\end{equation}
The thresholded recurrence plot can be expressed as follows:
\begin{equation}
    R_{m,n}=\Theta(\epsilon-D_{m,n}) ~~\&~~ m,n=1...N-(d-1)\tau
\end{equation}
Here, the threshold was determined based on the Fixed Amount of Neighbors (FAN) Scheme as follows:
\begin{equation}
    \epsilon_i = D_{i, (k)} \quad \text{where } k = \max(1, \text{round}(N \cdot RR))
\end{equation}
where $D_{m,n}, R_{m,n},~\Theta,~\epsilon,~||.||$, $N$ denotes the Distance matrix, Recurrence plot, Heaviside function, Threshold value, $L_2-$ norm, and total number of data points, respectively.

Although the FAN thresholding scheme results in an asymmetric recurrence plot ($R_{m,n} \neq R_{n,m}$), every row is ensured to have the same recurrence density. The recurrence plot consists of binary values, 0 and 1, represented by the white and colored points, respectively. The topology of the Recurrence Plot qualitatively reveals the system's nature: homogeneous point distributions denote stochasticity, diagonal lines indicate periodicity, and vertical or horizontal structures correspond to intermittency \cite{marwan2007recurrence}. Subsequently, certain recurrence parameters (RQA measures) were calculated to quantify the plot's topology, characterizing the underlying dynamics via the distribution of diagonal, vertical, and hornature structures on the resulting thresholded recurrence plots.  Table\ref{tab:RQA_FEATURES} explains the recurrence measures evaluated for each signal across regimes. In addition, three-dimensional phase plots were constructed to qualitatively observe the evolution of system dynamics.

\begin{table*}[!t]
\centering
\caption{The List of Recurrence Quantification Features}
\label{tab:RQA_FEATURES}
\fontsize{8pt}{10pt}\selectfont
\begin{tabular}{m{0.25\linewidth} >{\centering\arraybackslash}m{0.35\linewidth} m{0.35\linewidth}}
\toprule
\textbf{Parameter} & \textbf{Formula} & \textbf{Description} \\
\midrule

1. Recurrence Rate &
    $\displaystyle RR = \frac{1}{N_x N_y} \sum_{i,j} R_{i,j}$ &
    Represents the density of shared recurrence points (probability of recurrence). \\
\addlinespace

2. Determinism &
    $\displaystyle DET = \frac{\sum_{l \ge L_{\min}} l P(l)}{\sum_{i,j} R_{i,j}}$ &
    Proportion of recurrence points that form diagonal lines. \\
\addlinespace

3. Average Diagonal Length &
    $\displaystyle L = \frac{\sum_{l \ge L_{\min}} l P(l)}{\sum_{l \ge L_{\min}} P(l)}$ &
    The average length of diagonal lines on the recurrence plot. \\
\addlinespace

4. Longest Diagonal Line &
    $\displaystyle L_{\max} = \max(\{l_i\})$ &
    The maximum length of diagonal lines on the recurrence plot. \\
\addlinespace

5. Entropy &
    $\displaystyle ENTR = - \sum_{l \ge L_{\min}} p(l) \ln p(l)$ &
    The Shannon entropy of the probability distribution of diagonal line lengths. \\
\addlinespace

6. Laminarity [Vertical] &
    $\displaystyle LAM_V = \frac{\sum_{v \ge V_{\min}} v P(v)}{\sum_{i,j} R_{i,j}}$ &
    Proportion of recurrence points that form vertical lines. \\
\addlinespace

7. Laminarity [Horizontal] &
    $\displaystyle LAM_H = \frac{\sum_{h \ge H_{\min}} h P(h)}{\sum_{i,j} R_{i,j}}$ &
    Proportion of recurrence points that form horizontal lines. \\
\addlinespace

8. Trapping Time [Vertical] &
    $\displaystyle TT_V = \frac{\sum_{v \ge V_{\min}} v P(v)}{\sum_{v \ge V_{\min}} P(v)}$ &
    The average length of the vertical lines. \\
\addlinespace

9. Trapping Time [Horizontal] &
    $\displaystyle TT_H = \frac{\sum_{h \ge H_{\min}} h P(h)}{\sum_{h \ge H_{\min}} P(h)}$ &
    The average length of the horizontal lines. \\
\addlinespace
\addlinespace

10. Longest Vertical Line &
    $\displaystyle V_{\max} = \max(\{v_i\})$ &
    The length of the longest vertical line. \\
\addlinespace
\addlinespace

11. Longest Horizontal Line &
    $\displaystyle H_{\max} = \max(\{h_i\})$ &
    The length of the longest horizontal line. \\
\addlinespace
\bottomrule

\addlinespace
\multicolumn{3}{p{0.95\linewidth}}{\footnotesize \textbf{Notations:} \textbf{$R_{i,j}$}: Recurrence Matrix (1 if states $i, j$ recur, 0 otherwise); \textbf{$N_x, N_y$}: Total state vectors for signals X and Y; \textbf{$P(l), P(v), P(h)$}: Histograms of diagonal, vertical, and horizontal lines; \textbf{$L_{\min}, V_{\min}, H_{\min}$}: Minimum length thresholds; \textbf{$p(l)$}: Probability distribution of diagonal line lengths; \textbf{$L_{\max}, V_{\max}, H_{\max}$}: Longest diagonal, vertical, and horizontal lines.}
\end{tabular}
\end{table*}

\subsection{Statistical-Spectral Analysis}
Apart from the Recurrence quantification analysis, certain Statistical and Spectral features were computed for each segment of both signals, as described in Table \ref{tab:STAT_SPEC_FEATURES}. This included predominantly linear measures: Lag-1 Autocorrelation, Decorrelation Time, Harmonic Ratio, Zero-Crossing Rate, Dominant Frequency, Dominant Power Ratio, Spectral Centroid, Spectral Slope; along with a few other non-linear measures: Skewness, Kurtosis, Spectral Entropy. These attributes were computed for both signals and were integrated into another feature set (apart from RQA features) to enable comparative analysis. Furthermore, to elucidate the spectral characteristics, the temporal evolution of the frequency of various events was analyzed using Continuous Wavelet Transform (CWT), whereby for each segment, CWT was computed to generate a scalogram. These individual scalograms were examined to observe how the signal frequency content evolved.

\begin{table*}[!t]
\centering
\caption{The List of Statistical and Spectral Features}
\label{tab:STAT_SPEC_FEATURES}
\fontsize{8pt}{10pt}\selectfont
\begin{tabular}{m{0.24\linewidth} >{\centering\arraybackslash}m{0.30\linewidth} m{0.40\linewidth}}
\toprule
\textbf{Parameter} & \textbf{Formula} & \textbf{Description} \\
\midrule

\multicolumn{3}{c}{\textbf{Statistical Measures}} \\
\midrule
\addlinespace
1. Skewness (Third Standardized Moment) &
    $\displaystyle \frac{1}{N} \sum_{i=1}^{N} \left(\frac{x_i - \mu}{\sigma}\right)^3$ &
     Quantifies the asymmetry of the probability distribution about its mean. \\
\addlinespace

2. Kurtosis (Fourth Standardized Moment)&
    $\displaystyle \frac{1}{N} \sum_{i=1}^{N} \left(\frac{x_i - \mu}{\sigma}\right)^4$ &
     Quantifies the concentration of probability mass in the tails of the distribution. \\
\addlinespace

3. Lag-1 Autocorrelation Coefficient &
    $\displaystyle \rho_1 = \frac{E[(x_t - \mu)(x_{t+1} - \mu)]}{\sigma^2}$ &
    Quantifies the linear correlation between adjacent observations in the time series. \\
\addlinespace

4. Decorrelation Time &
    $\displaystyle \tau_{corr} = \min \{ \tau \mid \rho_\tau < 1/e \}$ &
    The time lag required for the autocorrelation function to decay to $1/e$ of its zero-lag value. \\
\addlinespace

5. Harmonic Ratio &
   $\displaystyle H_R = 10 \log_{10} \left( \frac{\rho_{max}}{1 - \rho_{max}} \right)$ &
   Estimates signal periodicity as the Harmonics-to-Noise Ratio (HNR) in decibels, derived from the maximum autocorrelation coefficient $\rho_{max}$. \\
\addlinespace

6. Zero-Crossing Rate &
    $\displaystyle \frac{1}{2N} \sum_{i=1}^{N-1} |\text{sgn}(x_{i+1}) - \text{sgn}(x_i)|$ &
    The fractional rate of algebraic sign changes in the signal sequence. \\

\midrule
\multicolumn{3}{c}{\textbf{Spectral Measures}} \\
\midrule

7. Dominant Frequency &
    $\displaystyle f_{dom} = \arg \max_f P(f)$ &
    The frequency component corresponding to the global maximum of the power spectral density. \\
\addlinespace

8. Dominant Power Ratio &
    $\displaystyle \frac{P(f_{dom})}{\sum P(f)}$ &
    The proportion of total spectral energy contained within the dominant frequency component. \\
\addlinespace

9. Spectral Centroid &
    $\displaystyle f_c = \frac{\sum f \cdot P(f)}{\sum P(f)}$ &
    The energy-weighted mean frequency represents the center of mass of the power spectrum. \\
\addlinespace

10. Spectral Entropy &
    $\displaystyle -\sum \frac{P(f)}{\sum P} \log_2 \left( \frac{P(f)}{\sum P} \right)$ &
    Quantifies the disorder or flatness of the spectral distribution (Shannon entropy of the normalized spectrum). \\
\addlinespace

11. Spectral Slope &
    $\displaystyle \beta \text{ where } P(f) \propto \frac{1}{f^\beta}$ &
    The rate of spectral energy decay with respect to frequency (on a logarithmic scale). \\
\addlinespace
\bottomrule

\addlinespace
\multicolumn{3}{p{0.95\linewidth}}{\footnotesize \textbf{Notations:} $N$: Signal length; $\mu$: Mean; $\sigma$: Standard Deviation; $\rho_\tau$: Autocorrelation at lag $\tau$; $\text{sgn}$: Signum function; $P(f)$: Power Spectral Density; $1/e \approx 0.367$.}
\end{tabular}
\end{table*}

\subsection{Stacking Ensemble Framework}
\begin{figure}[pos=H]
    \centering
    \includegraphics[width=\linewidth]{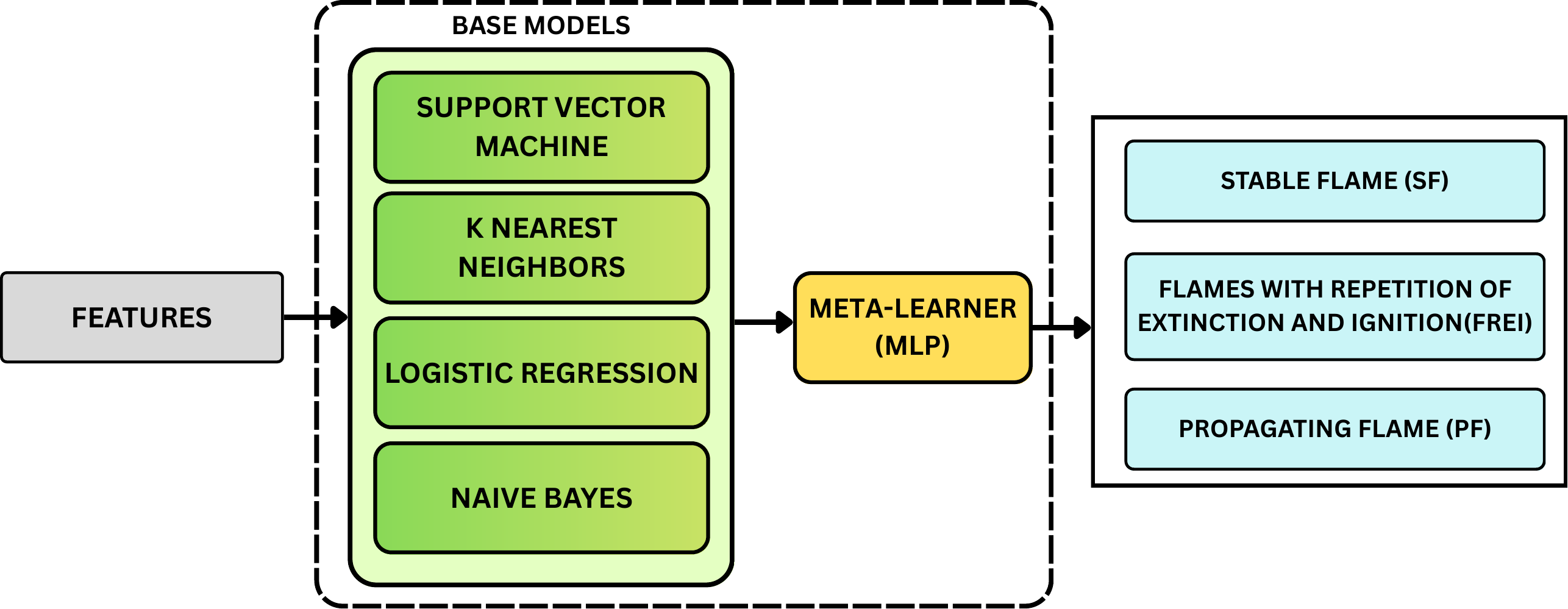}
    \caption{Stacking ensemble Framework for Mesoscale Flame regime Classification, describing the usage of a comprehensive suite of four classifiers followed by a meta-learner.}
    \label{Classification}
\end{figure}

The stacking ensemble framework for ternary flame regime classification implemented has been illustrated in Figure \ref{Classification}. A comprehensive suite of four base classifiers, namely Support Vector Machine \cite{cortes1995support}, K Nearest Neighbors \cite{cover1967nearest}, Naive Bayes  \cite{domingos1997optimality}, and Logistic Regression  \cite{cox1958regression}, was employed to facilitate a comparative evaluation. Rigorous hyperparameter analysis and model optimization were performed using GridsearchCV with stratified five-fold cross-validation, systematically tuning each classifier to maximize the weighted F1-score. Each feature was Z-score normalized before training to ensure scale invariance, followed by Synthetic Minority Over-sampling Technique (SMOTE)  \cite{chawla2002smote} to mitigate the class imbalance. Restricting these preprocessing steps to the training set effectively prevented data leakage during validation. In order to ensure robust testing, signals from unseen operating conditions within each class were reserved exclusively as held-out test set.

Finally, the stacking ensemble aggregated the optimized base classifiers' decision boundaries. Meta-features were generated by extracting out-of-fold probabilistic predictions using stratified five-fold cross-validation, a strategy strictly employed to prevent data leakage during the meta-training phase. These aggregated probabilities served as the input features for a Multi-Layer Perceptron (MLP) meta-learner, which was trained to learn complex, non-linear combinations of the base models' outputs. The final ensemble performance was evaluated on the independent test set using standard classification metrics and confusion matrices.

Flame regime classification was performed with two feature sets independently. Firstly, based on the recurrence quantification analysis (RQA) features, followed by Statistical-Spectral measures (each feature family, including the combined measures derived from both signals), to facilitate a comparative analysis. All the above-mentioned comprehensive model optimization and training procedures were independently applied to both feature sets. The predictive performance was evaluated using confusion matrices, ROC curves, and standard performance metrics (accuracy, precision, recall, and F1 scores). The analysis was further extended to investigate the class separability through linear and non-linear dimensionality reduction maps, namely Principal Component Analysis (PCA) and Isomap (Isometric Mapping), with decision boundaries plotted on them using a Random-Forest classifier for better interpretability of their class separation. PCA \cite{hotelling1933analysis} linearly projects high-dimensional data onto directions of maximum variance via matrix factorization, whereas Isomap \cite{tenenbaum2000isomap} unrolls curved, non-linear manifolds by calculating geodesic distances along neighbor graphs.

\section{Results}
The results section has been structured as follows. Firstly, we decode the Flame regime dynamics using recurrence plots, phase diagrams, and scalograms. This would be followed by the stacking ensemble classification based on the independent feature sets: RQA and Statistical-Spectral measures.

\subsection{Flame Regime Dynamics Characterization}
In this section, we describe the different flame regimes and their associated key global characteristics. The premixed fuel-air mixture enters the combustor tube at $x=0$ and continuously gains heat from the combustor walls as it moves downstream. Near the section of the combustor directly above the external heater, the mixture auto-ignites, and the resulting flame tends to traverse upstream against the incoming reactants. Depending on the operating conditions, three distinct flame behaviors were observed: i) Stable Flame (SF), ii) Flames with Repetitive Extinction and Ignition (FREI), and iii) Propagating Flame (PF). Figure \ref{fig:Regime_Map}  shows the regime map obtained at different $Re$ and $\Phi$, depicting the occurrence of different flame regimes as a function of operating conditions, with these regimes being highlighted with different colors. The physics behind these observations for each flame regime will be discussed in detail in the following subsections.

In essence, in the Stable Flame (SF) regime, the upstream-traversing flame stabilizes at a characteristic location and remains stationary. In the Flames with Repetitive Extinction and Ignition (FREI) regime, the auto-ignited flame traverses upstream and extinguishes after traveling a characteristic distance; the mixture auto-ignites again after a time delay, and the flame repeats in cycles of ignition, travel, and extinction. In the Propagating Flame (PF) regime, the auto-ignited flame traverses the entire length of the combustor tube and is only extinguished at the upstream meshed constriction. After a characteristic delay, the mixture auto-ignites again, and a new propagating flame starts to traverse towards the upstream end of the tube. While the FREI regime was encountered under stoichiometric and fuel-rich conditions, PF was observed in fuel-lean conditions. Interestingly, propagating flame were found to exhibit thermo-acoustic instabilities during their propagation phase, where the heat release rate and pressure fluctuations coupled at frequencies close to the natural harmonic of the combustor tube \cite{aravind2023dynamics}.

\subsubsection{Stable Flame (SF)}

\begin{figure}[pos=H]
    \centering
    \includegraphics[width=\linewidth]{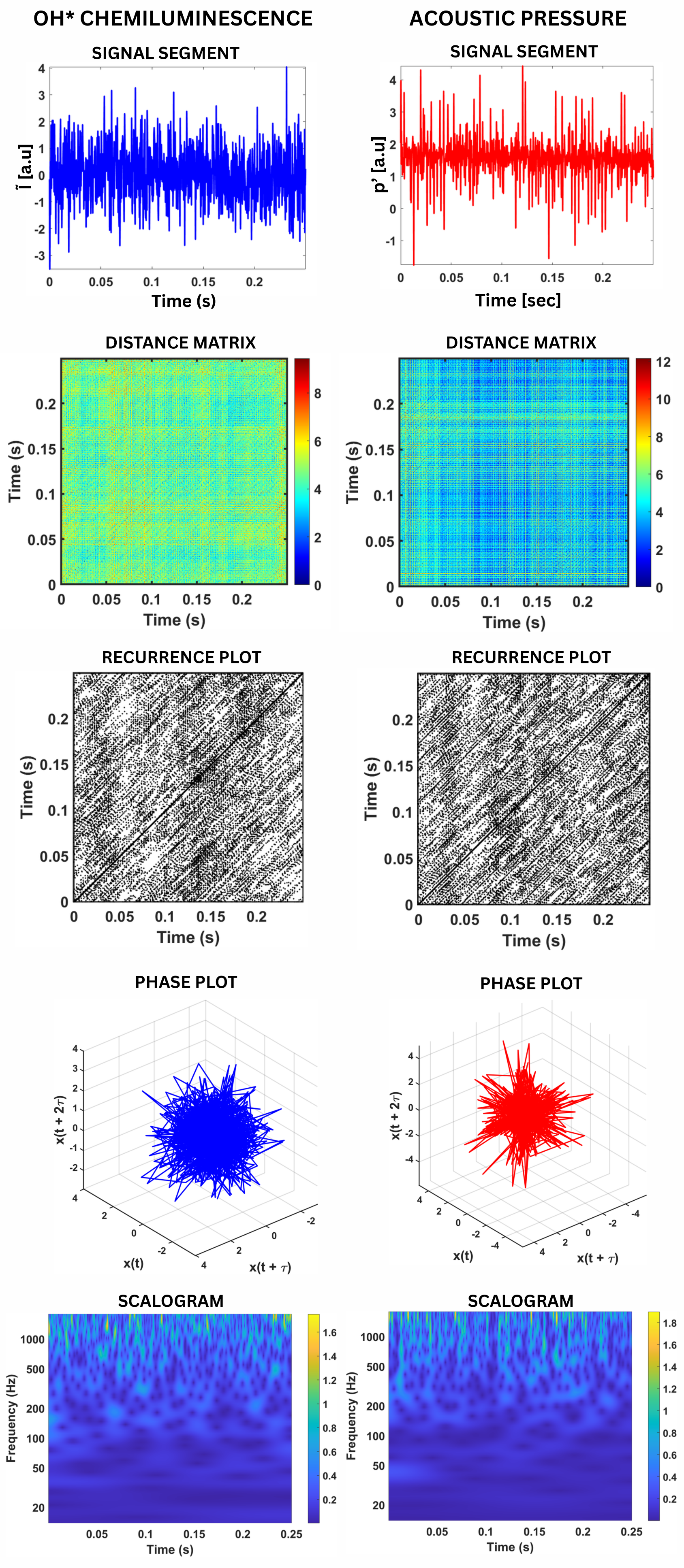}
    \caption{Plots depicting the Dynamical signal characteristics of the Stable Flame (SF) regime observed at $\Phi=1.0$ and $Re=224$. The left and the right columns display the signal segment and the plots corresponding to OH* chemiluminescence and Acoustic pressure signals, respectively.}
    \label{SF}
\end{figure}

Figure \ref{SF} clearly elucidates the dynamical signatures of the Stable Flame regime (observed \(\Phi = 1.0\) and \(Re = 224\)). The left and right panels describe the characteristics associated with instantaneous OH\(^*\) Chemiluminescence intensity and Acoustic pressure signals, respectively. The chosen signal segments have been displayed on the top, followed by their corresponding recurrence distance matrices alongside their corresponding thresholded recurrence plots in the middle, and finally, the three-dimensional phase plots and Scalograms (obtained using Continuous Wavelet Transform) at the bottom. The actual intensity signals \(\tilde{I}\) showed fluctuations around a non-zero mean value (\(\sim O(1)\)), which is related to the stable flame's mean heat release. However, since these signals were Z-scale normalized, the plots here describe their variation about the mean heat release rate. The features of the stable flame regime closely resemble high-frequency noise. Also, the acoustic pressure emission from the stable flame was observed to exhibit stochastic oscillations. The associated scalograms with both signals also revealed uniformly distributed amplitude levels, predominantly exhibiting high-frequency noise. Furthermore, the recurrence plot in panels displays features characteristic of stochasticity and non-periodic behavior, with homogeneous discontinuous points distribution \cite{mohan2023self, marwan2007recurrence}, which are attributed to high-frequency noise, with the major diagonal line of these plots being representative of their self-recurrence. Analogous plot characteristics were observed for both OH\(^*\) chemiluminescence intensity and acoustic pressure signals. Furthermore, the phase plots for both signals demonstrated stochastic trajectories typically signifying the presence of high-frequency stochastic noise.

\subsubsection{Flames with Repetitive Extinction and Ignition (FREI)}
\begin{figure}[pos=H]
    \centering
    \includegraphics[width=\linewidth]{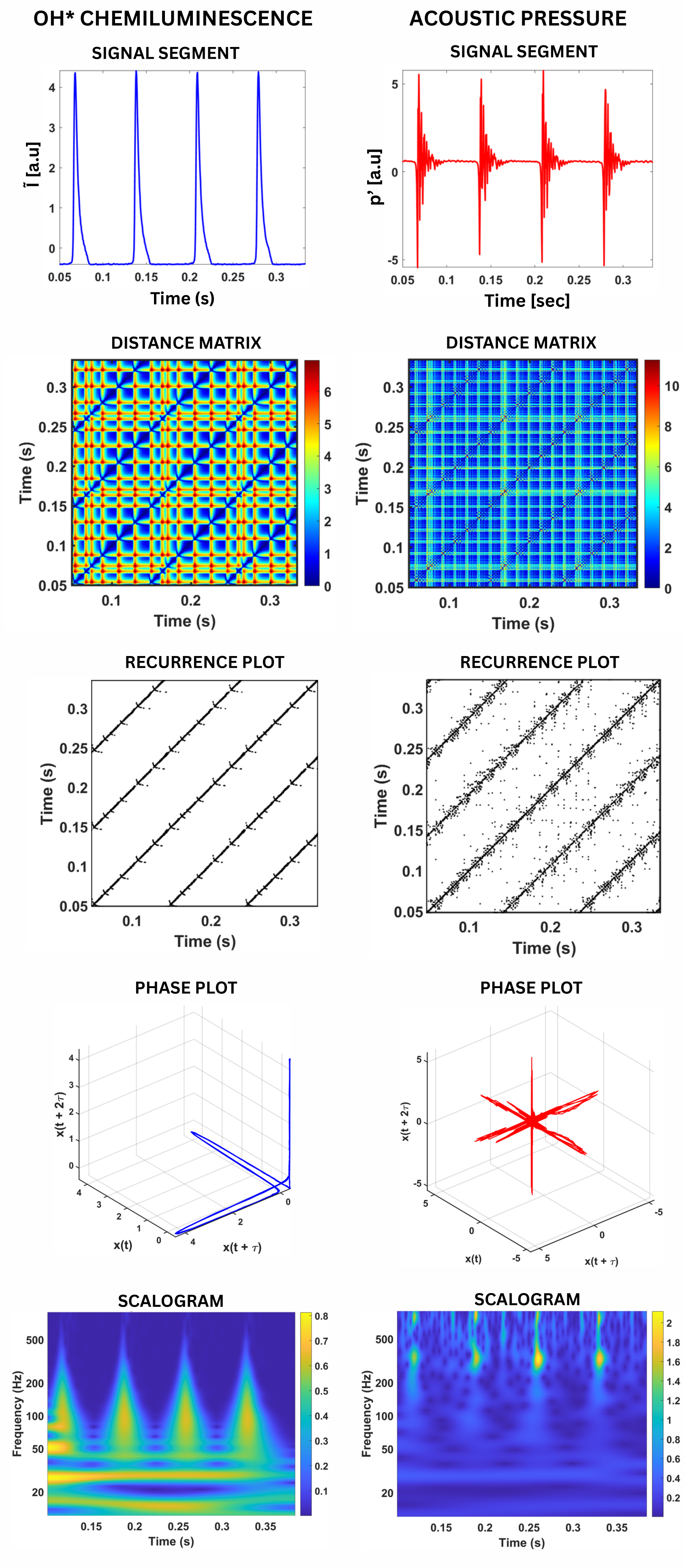}
    \caption{Plots depicting the Dynamical signal characteristics of the Flames with Repetitive Extinction and Ignition (FREI) regime observed at $\Phi=1.0$ and $Re=96$. The left and the right columns display the signal segment and the plots corresponding to OH* chemiluminescence and Acoustic pressure signals, respectively.}
    \label{FREI}
\end{figure}

Figure \ref{FREI} presents the corresponding dynamical signatures of the FREI regime observed at $\Phi = 1.0$ and $Re = 96$. The time series of the normalized heat release rate, $\tilde{I}$, exhibits a periodic rise and decay. The rise corresponds to flame ignition, while the subsequent decay represents the upstream propagation phase that terminates in extinction after a characteristic travel distance, at which point the heat release rate (scaled with $\tilde{I}$) drops to zero. A concurrent peak is observed in the acoustic pressure signal ($p^{\prime}$) at ignition, corresponding to the explosive nature of the ignition event.  The corresponding scalogram of $\tilde{I}$ shows a distributed peak, reflecting the impulsive nature of the initial heat release. In addition, a dominant low-frequency mode is observed, which is attributed to the periodic ignition-extinction cycle characteristic of the FREI regime. This ignition-extinction cycle frequency can be related to the mixture convection time scale, reaction time scale, and flame propagation speed \cite{aravind2023dynamics}. A similar broadband frequency content is observed in the acoustic pressure signal ($p^{\prime}$) during the ignition phase. 

The scalogram further indicates that the ignition event excites a wide range of frequencies, with the first natural acoustic mode of the duct being the most prominent. However, following ignition, the excited pressure oscillations decay and the system returns to a quiescent state. Excitation of the natural harmonic does not lead to sustained pressure oscillations. The underlying reason for this behavior can be understood from the flame location during the FREI cycle. The flame ignites near the downstream end of the duct owing to external heating and subsequently propagates upstream. However, the flame travel distance is nearly an order of magnitude smaller than the duct length, and consequently, the flame remains confined to the downstream region of the duct. In this region, coupling between the unsteady heat release rate and the acoustic field is unfavorable for the growth of thermoacoustic oscillations (approximating the duct to have a closed upstream end and an open downstream end) \cite{carvalho1989definition}. A similar observation was recently reported in an annular combustor by Mohan and Mariappan \cite{mohan2023self}. In contrast, when the flame propagates further upstream beyond the half-length, as in propagating flame regimes, strong thermoacoustic feedback can develop, leading to sustained acoustic oscillations.

The recurrence plots of the heat release rate display features characteristic of relaxation oscillators with mild additive noise. In contrast, the acoustic pressure signal following ignition exhibits features resembling the impulse response of an overdamped oscillator. The periodicity of both $\tilde{I}$ and $p^{\prime}$ is manifested as parallel diagonal lines in the recurrence plots, with the spacing between diagonals corresponding to their characteristic repetition time period. The phase portrait of $\tilde{I}$ forms a closed, spiraling loop indicative of deterministic periodic dynamics typical of a relaxation oscillator, whereas the phase portrait of $p^{\prime}$ revealed a closed star-shaped limit cycle characterizing their periodic pattern with amplitude modulation resembling a damped oscillator.

\subsubsection{Propagating Flame (PF)}
\begin{figure}[pos=H]
    \centering
    \includegraphics[width=\linewidth]{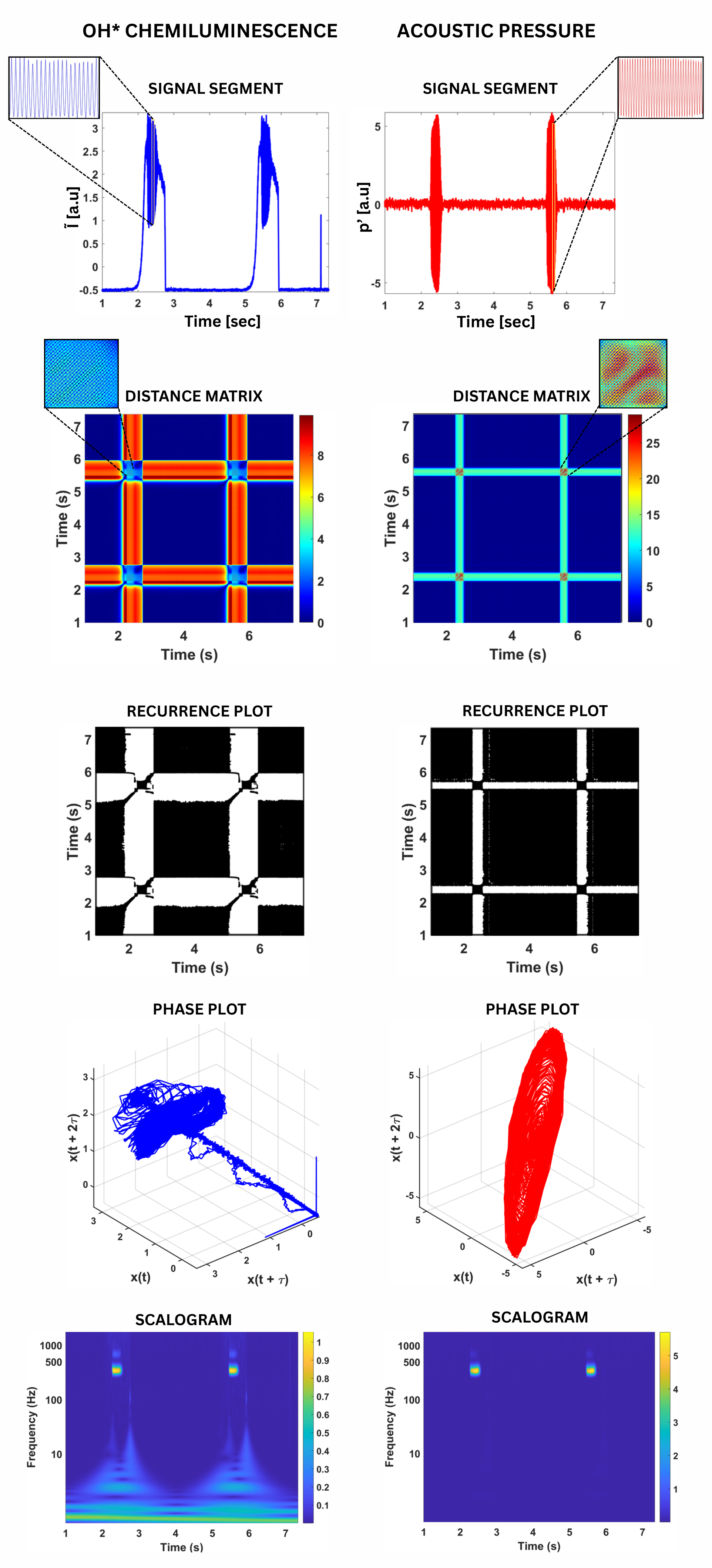}
    \caption{Plots depicting the Dynamical signal characteristics of the Propagating Flame (PF) Regime observed at $\Phi=0.8$ and $Re=64$. The left and the right columns display the signal segment and the plots corresponding to OH* chemiluminescence and Acoustic pressure signals, respectively.}
    \label{PF}
\end{figure}

Figure \ref{PF} illustrates the plots describing the dynamics associated with the propagating flame (PF) regime, observed at operating conditions $\Phi = 0.8$ and $Re = 64$. Unlike the FREI regime discussed in the previous subsection, the acoustic emission from a propagating flame is significantly stronger during the propagation phase than during the ignition event, owing to the presence of thermoacoustic coupling. Relatively large-amplitude oscillations are observed as the flame propagates into the upstream half of the combustor tube \cite{aravind2023dynamics}, and these fluctuations are reflected as corresponding oscillations in $\tilde{I}$. As the flame reaches the upstream end of the combustor, it is extinguished by the meshed constriction, at which point, the heat release rate decays to zero. The cycle then repeats when the unburnt reactant mixture convects downstream, reaches the ignition location, and triggers the subsequent ignition event.
The scalogram of $\tilde{I}$, similar to that observed for the FREI regime, indicates a distributed energy spectrum and its decay during the initial phase following ignition, when the flame is located near the downstream end of the channel, where thermoacoustic coupling is unfavorable. As the flame subsequently propagates upstream, a monotone response emerges, corresponding to the first natural harmonic of the combustor tube. A similar single-frequency response is observed in the scalogram of the acoustic pressure signal ($p^{\prime}$). The oscillations in $\tilde{I}$ and $p^{\prime}$ are coupled, and the coupling frequency can be estimated using frequency-domain analysis of the combustor tube by assuming a closed boundary at the upstream end and an open boundary at the downstream end \cite{aravind2023dynamics}.

The recurrence plots of both signals display large square patches, which correspond to stochastic noise between successive propagating flame cycles. The dynamics associated with thermoacoustic oscillations during the propagation phase appear as closely spaced diagonal lines, with the characteristic time scale given by the spacing between these diagonals, as highlighted in the distance matrices. This spacing corresponds to the inverse of the thermoacoustic coupling frequency. It should be noted that the intermediate transient dynamics associated with ignition, propagation, and extinction are not clearly discernible in the recurrence plots, as thresholding effects may merge the narrow temporal gaps associated with high-frequency thermoacoustic oscillations or obscure certain features associated with ignition and extinction. Overall, these patterns exhibit characteristics typical of intermittent signals. The phase space representation of the acoustic pressure signal exhibits a disc-like structure, which can be attributed to the transient growth and decay of the oscillations. In contrast, the phase space of $\tilde{I}$ exhibits a structure resembling a mace, consisting of a shaft and a head. The shaft is associated with the ignition and extinction events, while the head corresponds to the thermoacoustic oscillations during the flame propagation phase.

\subsection{Flame Regime Classification}
Following the Recurrence and Statistical-Spectral analyses performed based on OH* Chemiluminescence and acoustic signals, the resulting feature sets computed were integrated to create two comprehensive families of features (Recurrence and Statistical-Spectral attributes/features). These two feature sets were independently fed into the stacking ensemble framework to clearly distinguish the unique dynamic signatures of the flame regime observed in mesoscale combustors. Furthermore, dimensionality reduction maps (Isomap and PCA) were utilized to visualize these feature sets onto two-dimensional latent spaces. The prediction results based on these feature sets have been explained in detail.

\subsubsection{Prediction based on Recurrence Quantification Features}
\begin{figure}[pos=H]
    \centering
    \begin{subfigure}{\linewidth}
        \centering
        \includegraphics[width=\linewidth]{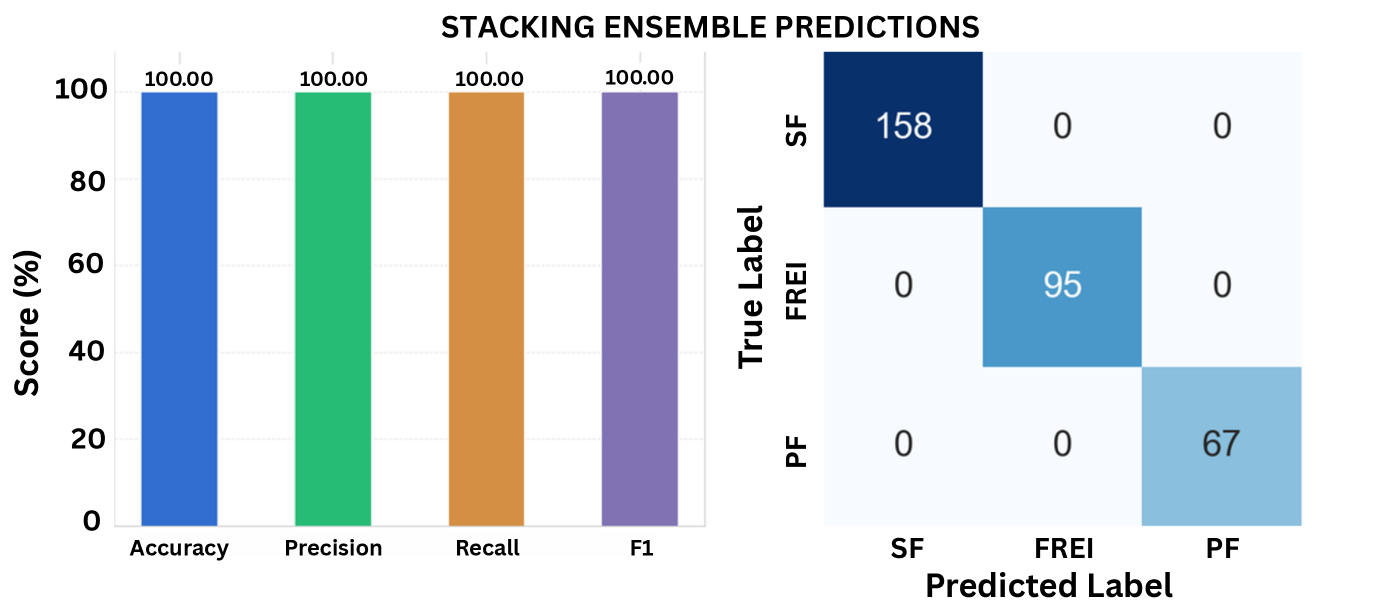}
        \label{fig:ENSEMBLE_RQA}
    \end{subfigure}
    \begin{subfigure}{0.88\linewidth}
        \centering
        \includegraphics[width=\linewidth]{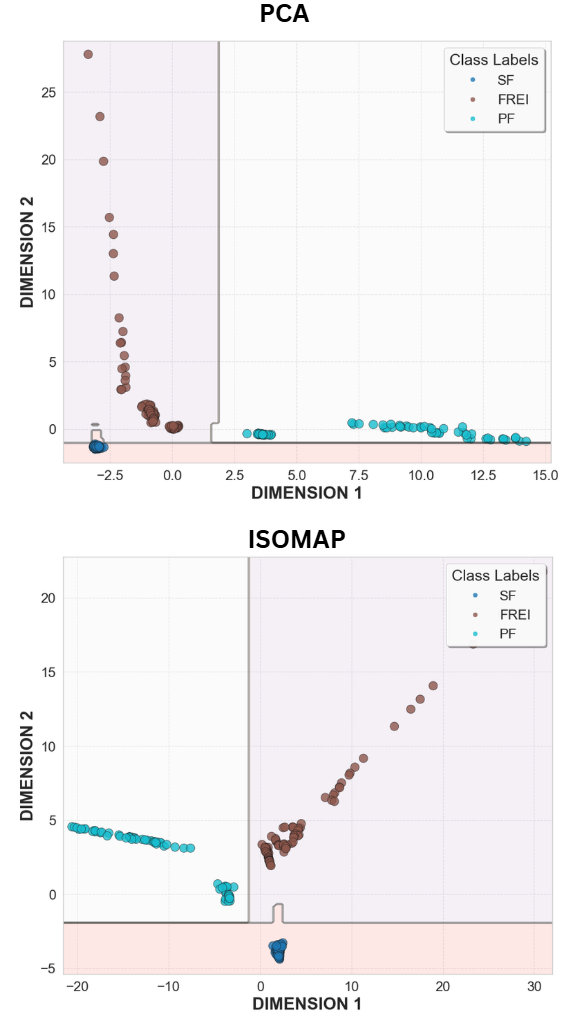}
        \label{fig:Isomap_PCA_RQA}
    \end{subfigure}
    \caption{Predictions based on RQA measures from Stacking ensemble Framework. Confusion Chart, Performance metrics, and the Dimensionality Reduction Maps (PCA, Isomap) based on RQA features have been presented here.}
    \label{RQA_CLASSIFICATION}
\end{figure}

The results based on the classification using a combined feature set comprising 20 Recurrence quantification measures (excluding recurrence rate, since the analysis was performed at a fixed recurrence rate) have been depicted in Figure \ref{RQA_CLASSIFICATION}. The confusion chart and performance metrics display highly accurate predictions with accuracy, precision, recall, and F1 score. Both the dimensionality reduction maps (Isomap, PCA) show inherent clustering as feature clouds corresponding to the three flame regimes. This observation stands as concrete proof to support the prediction accuracy observed from the stacking ensemble framework. Hence, this feature set, which consisted of non-linear RQA measures, was able to clearly distinguish the flame regimes based on their unique dynamic signatures observed.

\subsubsection{Prediction based on Statistical-Spectral Features}
\begin{figure}[pos=H]
    \centering
    \begin{subfigure}{\linewidth}
        \centering
        \includegraphics[width=\linewidth]{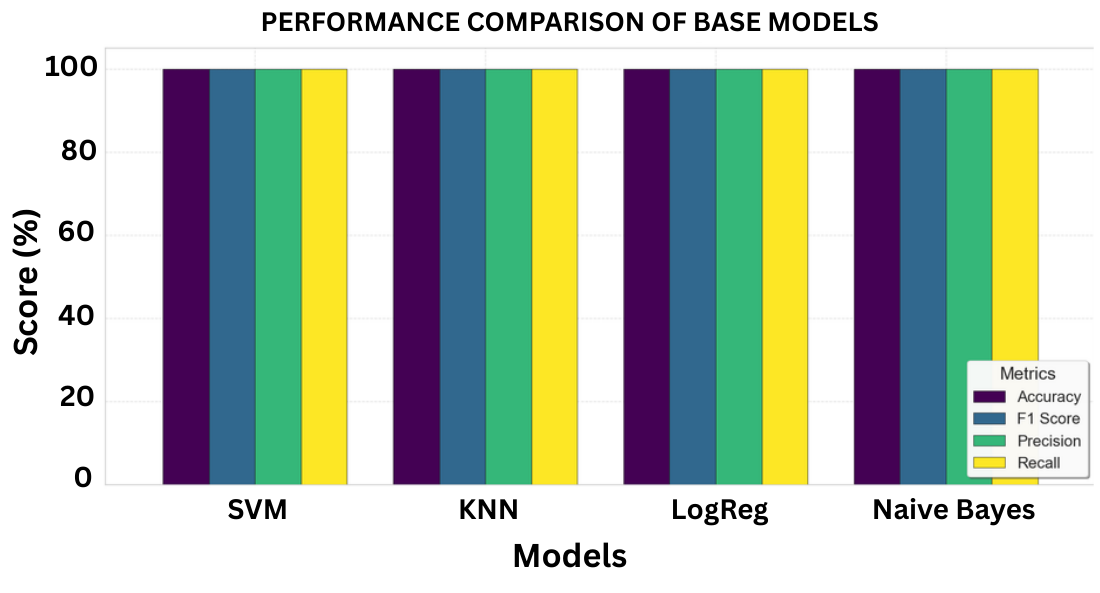}
        \label{fig:ENSEMBLE_SPEC}
    \end{subfigure}
    \begin{subfigure}{0.88\linewidth}
        \centering
        \includegraphics[width=\linewidth]{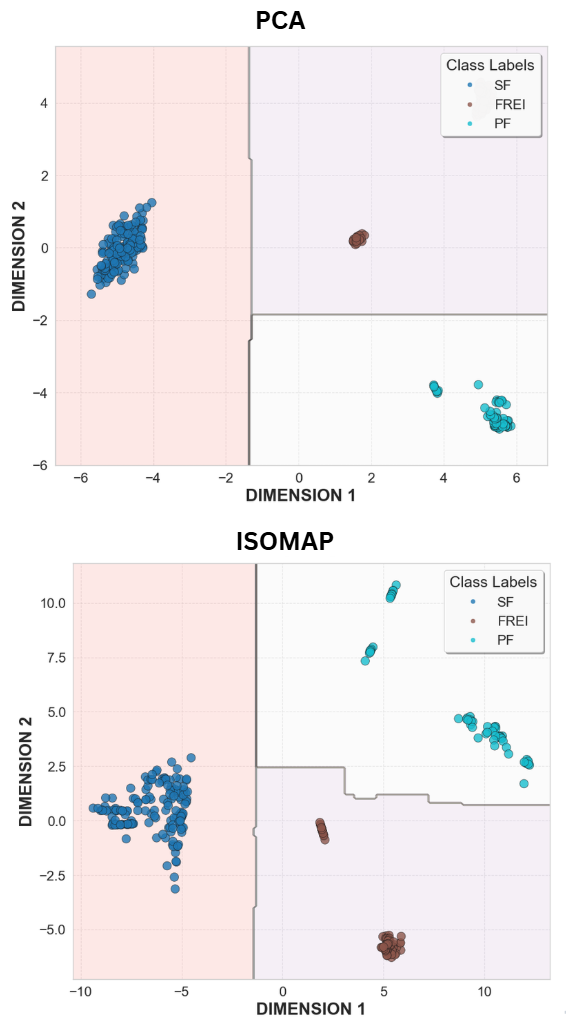}
        \caption{Dimensionality Reduction Maps }
        \label{fig:Isomap_PCA_SPEC}
    \end{subfigure}
    \caption{Predictions based on Statistical-Spectral measures illustrating base models' performance, followed by the Dimensionality reduction maps (PCA, Isomap) based on Statistical-Spectral features have been presented here.}
    \label{STAT_SPEC_CLASSIFICATION}
\end{figure}

The feature set, comprising 22 statistical-spectral measures evaluated independently for both signals, when combined and provided as input to the stacking ensemble framework, yielded the predictions shown in Fig. \ref{STAT_SPEC_CLASSIFICATION}. The performance metrics of the individual base models have been presented along with the dimensionality reduction maps, which clearly demonstrate that the selected attributes are both sufficient and capable of capturing the distinct dynamical signatures of all three flame regimes. Consequently, the meta-learner predictions were synonymous with those observed for RQA Features. This is an interesting and seemingly useful observation, since these (predominantly linear) measures have much lower computational cost than the recurrence-based feature calculations. Nevertheless, recurrence-based analysis provided profound insights into the nonlinear dynamic behavior, elegantly based on structural patterns in recurrence plots. However, from a prediction standpoint, the chosen Statistical-Spectral features were sufficient to clearly identify the three different flame regimes flawlessly.

\section{Discussion}
Phase space reconstruction, using the selected time delay and embedding dimension, along with the corresponding recurrence plots (RP), was employed to understand the characteristics of the different flame regimes in the mesoscale combustor. In addition to the recurrence analysis based on the two univariate signals discussed above, cross-recurrence analysis (CRQA) was also performed\footnote{The cross-recurrence analysis has been presented with the Supplementary Material for interested readers.}. However, to address the present objective of flame regime classification, only the individual recurrence-based features were considered, as they were sufficient and encompassed the features associated with both the time-series signals. A potential limitation of recurrence-based analysis is that thresholding may fail to capture certain fine-grained details of the distance matrix, particularly when multiple time scales that are widely separated are present, such as the thermoacoustic coupling time scale and the ignition-extinction cycle time scale in propagating flame.

The features computed by Statistical-Spectral analysis were specifically selected to characterize Z-scaled signals, synthesizing distributional, frequency-based, and temporal descriptors to capture the multi-scale dynamics of the underlying system. The skewness and kurtosis characterize the impulsiveness (tail-heaviness) and asymmetry of the signal. The dominant frequency, spectral centroid, harmonic ratio, and the dominant spectral power describe the primary periodicity, spectral center of mass, and energy concentration. Furthermore, the spectral slope and spectral entropy quantify the disorder and decay rate associated with the power spectrum, while the zero crossing rate additionally captures the oscillation frequency and noisiness. The autocorrelation lag and decorrelation time describe the temporal memory and deterministic predictability. Hence, these attributes, computed from the experimental signals, showcased precise predictions from the stacking ensemble framework. Consequently, this comprehensive multi-domain characterization provides a robust discriminative basis for flame regime classification, effectively distinguishing stochasticity, periodicity, and intermittency. 

In this study, we employed SVM (maximizing the separating hyperplane margin), KNN (based on Euclidean proximity), Logistic Regression (sigmoid-based probabilistic modeling), and Naïve Bayes (conditional likelihood estimation), followed by an MLP-based meta-learner to stack these predictions. The observed near-perfect classification accuracy on the held-out test set reflects the fundamental physical distinctness of the identified regimes and does not result from typical statistical concerns such as model overfitting or data leakage. As clearly demonstrated by the dimensionality reduction maps, the three flame regimes appeared as well-separable manifolds in the chosen feature spaces, confirming that the selected features possessed the necessary discriminative power to resolve the system dynamics without ambiguity. Consequently, the task of the classifier was simplified to establishing this decision boundary, hence validating that both the recurrence-based and statistical-spectral feature sets served as robust and physically grounded representations of the underlying phenomena. Although individual base models yielded promising predictions, an ensemble learning approach was employed and prioritized for its algorithmic robustness. This can be supported by the argument that prediction results across various classifiers, along with dimensionality reduction maps, confirm that the exact predictions arise from the fundamental separability of the physical regimes observed using the chosen attributes, rather than from clustering artifacts specific to any given algorithm.

\section{Conclusion} 
In this article, the various flame regimes observed in mesoscale combustors are examined using both linear and nonlinear time-series analysis techniques. By varying the equivalence ratio and the mixture Reynolds number, OH$^*$ chemiluminescence and Acoustic pressure time-series data were acquired experimentally, along with high-speed flame imaging. A range of time-series analysis tools—including statistical and spectral measures, wavelet scalograms, phase-space reconstruction, and recurrence plots — were applied to elucidate the underlying flame dynamics.

Three distinct flame regimes were investigated in detail: Stable flame (SF), Flames with repetitive extinction and ignition (FREI), and Propagating flame (PF). Stable flame was found to stabilize at characteristic axial locations within the combustor and exhibited features closely resembling stochastic signals. In the FREI regime, the flame undergoes cyclic sequences of ignition, upstream propagation, and extinction. The heat release rate signal in this regime displays characteristics of a relaxation oscillator, whereas the corresponding acoustic pressure signal resembles the response of a damped oscillator over an ignition-extinction cycle. Spectrogram analysis reveals broadband frequency excitation during the ignition phase, followed by decay to quiescent conditions during the propagation phase that terminates in extinction. Although the energy spectrum is distributed, acoustic excitation is most prominent at the natural harmonic of the combustor immediately following ignition. However, due to the absence of sustained thermoacoustic feedback, persistent pressure oscillations are not observed. The recurrence plots exhibit diagonal line structures representative of the periodically repeating ignition-extinction cycles. In the propagating flame regime, ignition is followed by sustained upstream flame propagation toward the meshed upstream end of the combustor, where the flame is forced to extinguish. A new propagating flame subsequently forms when the premixed reactants convect downstream and re-ignite at the ignition location. As the flame propagates beyond approximately half the duct length toward the upstream mesh, pronounced thermoacoustic oscillations develop. During this phase, the scalograms of both the heat release rate and acoustic pressure signals exhibit a coupled, single-frequency response. However, the time scales associated with this thermoacoustic coupling phase are orders of magnitude smaller than the repetitive ignition--extinction time scale characteristic of propagating flame. As a result, the recurrence plots display large box-like structures corresponding to the propagating flame cycle, which overshadow the higher-frequency thermoacoustic oscillations during propagation, giving rise to signatures characteristic of intermittent dynamics.

Subsequently, the two feature sets constructed using Recurrence quantification and Statistical-Spectral measures (of both the signals) were fed independently into our stacking ensemble framework, which comprised a suite of four machine learning classifiers followed by an MLP-based meta-learner to observe the unique dynamical signatures corresponding to each of these flame regimes. It was observed that both the recurrence and Statistical-Spectral feature sets demonstrated exceptional discriminative power and inherent class separability, evidenced by the prediction plots (Confusion matrix and Performance chart), further visually supported by the dimensionality reduction maps, which confirm the consistent high accuracy across diverse classifiers. The recurrence-based study highlighted the core physical insights and their understanding. Nevertheless, for identifying the three regimes distinctly, the Statistical and Spectral set of features was sufficient and capable. Therefore, this comprehensive comparative analysis helped to elucidate the unique dynamics observed and to classify them based on these observed characteristics of each flame regime. Overall, this study enabled the understanding and characterization of the different flame regimes observed in mesoscale combustors by our two-pronged approach using linear and non-linear analysis, and subsequently an ensemble flame regime classification based on these acquired features.

\section*{Acknowledgments}
S.B. acknowledges support from the INAE (Indian National Academy of Engineering) Chair Professorship. A.A. acknowledges support from PMRF (Prime Minister Research Fellowship). The authors would like to express their gratitude to Mr. Arghya Paul from the Indian Institute of Science for his contributions to this study. 

\section*{Declaration of competing interest }
The authors report no conflict of interest. 

\section*{Supplementary Material}
The probability distributions comparing the dynamics of the signal segment with the entire time-series and the Cross-recurrence analysis have been provided in the Supplementary material. 
\textbf{Code availability:} The code supporting this study is available from the authors upon reasonable request.

\printcredits
\bibliographystyle{cas-model2-names}
\bibliography{cas-refs}

\end{document}